\documentstyle [12pt] {article}
\begin{document}
\thispagestyle{empty}
\begin{flushright} DESY 98-083\\July 1998\
\end{flushright}
\vspace{0.5in}
\begin{center}
{\Large \bf Naturally light sterile neutrinos \\[0pt]}
\vspace{1.0in}
{\sc Utpal Sarkar$^{(a,b)}$}\footnote{E-mail: utpal@prl.ernet.in}  

\vskip 0.8cm

\begin{small} 
$^{(a)}$Theory Group, DESY, Notkestra{\ss}e 85, 22603 Hamburg, Germany\\
$^{(b)}$Theory Group, Physical Research Laboratory,
Ahmedabad, 380 009, India \footnote{permanent address}\\
\end{small}
\end{center}

\vskip 2cm

\begin{abstract}

A simple  model  to accomodate light   sterile neutrinos  naturally 
with large mixing with  the usual neutrinos has been proposed. 
The standard model gauge group is extended  to include an $SU(2)_S$
gauge symmetry.  Heavy triplet higgs scalars  give small masses to the
left-handed neutrinos, while a heavy doublet  higgs scalar give mixing
with the  sterile  neutrinos of  the same  order  of magnitude. The neutrino
mass matrix thus obtained can explain the solar neutrino deficit, the
atmospheric neutrino deficit, the LSND data and hot dark matter. Lepton
number is violated  here  through decays of  the  heavy triplet higgs,
which generates the  lepton asymmetry of  the universe,  which in turn
generates a baryon asymmetry of the universe. 

\end{abstract}

\newpage 
\baselineskip 18pt 

Recently the Super Kamiokande \cite{atm}
announced a positive evidence of neutrino
oscillations. They attribute the $\nu_\mu$ deficit in the atmospheric 
neutrino to a $\nu_\mu$ oscillating into a $\nu_{atm}$, where 
$\nu_{atm}$ could be a $\nu_\tau$ or a
sterile neutrinos (which is a singlet under the standard model) with
$\Delta m_{atmos}^2 = m_{\nu_\mu}^2 - m_{atm}^2 \sim (0.5 - 6) \times
10^{-3} {\rm eV}^2 .$
There are also indications of neutrino oscillations in the neutrinos
coming from the sun. The solar neutrino deficit can be explained if one
considers $\nu_e \to \nu_{sol}$ oscillations 
(where $\nu_{sol}$ could be $\nu_\mu$ or $\nu_\tau$ or a sterile
neutrino) with the mass squared difference \cite{sol}
$\Delta m_{solar}^2 = m_{\nu_{sol}}^2 - m_{\nu_e}^2 \sim (0.3 - 1.2)
\times 10^{-5} {\rm eV}^2 ,$
This mass squared difference 
is for resonant oscillation \cite{msw}. If one assumes vacuum 
oscillation solution of the solar neutrino deficit, 
then this number will be several orders of magnitude 
smaller. If we assume a three generation 
scenario, $\nu_{atm}$ is then identified with $\nu_\tau$ and $\nu_{sol}$
could be a $\nu_\mu$ or a $\nu_\tau$. 
Consider, $\nu_{sol} \equiv \nu_\tau$, which implies,
$m_{\nu_\mu}^2 - m_{\nu_e}^2 = [m_{\nu_\mu}^2 - m_{\nu_\tau}^2] +
[m_{\nu_\tau}^2 - m_{\nu_e}^2 ] = \Delta m_{atmos}^2 + \Delta m_{solar}^2 
\sim 10^{-2} {\rm eV}^2 .$
Then we cannot explain the LSND result \cite{lsnd}, which 
announced a positive evidence of $\overline{\nu_\mu} \to 
\overline{\nu_e}$ oscillations with the mass squared difference 
(alternate explanation is not possible \cite{mann}) 
$\Delta m_{LSND}^2 = m_{\nu_\mu}^2 - m_{\nu_e}^2 \sim (0.2 - 2) 
{\rm eV}^2$. This conclusion is true even if we consider 
$\nu_{sol} \equiv \nu_\mu$.

As a solution to this problem one can say that either LSND result
is wrong or there has to be some other explanation for the solar
neutrino deficit. But a more popular solution is to extend the
standard model to incorporate a sterile neutrino and explain all
these experiments  \cite{rev}. 
Since LEP data \cite{lep} ruled out any possibility of
a fourth $SU(2)_L$ doublet left-handed neutrino, this fourth neutrino
has to be a sterile neutrino, which does not interact through any of the
standard model gauge bosons. Incorporating such light sterile neutrino with
large mixing with the other light neutrinos in
extensions of the standard model is non-trivial \cite{rev}.
Recently there is
one attempt to extend the radiative neutrino mass generation model by Zee
\cite{zee} to incorporate a sterile neutrino \cite{ma1}. 
However, that model cannot explain the baryon asymmetry of the universe. 

In this article, we propose a new scenario with an $SU(2)$ symmetry, 
which can provide a naturally light sterile
neutrino with large mixing with the other left-handed neutrinos. The
neutrino mass matrix with the sterile neutrinos can now explain the 
LSND data, the solar neutrino problem, the atmospheric neutrino anomaly
and the dark matter problem. The lepton number violation at a very
high scale generates a lepton asymmetry of the universe, which then
gets converted to the baryon asymmetry of the universe during the 
electroweak phase transition. 

We work in an extension of the standard model which 
includes a couple of heavy triplet higgs scalars \cite{triplet,ma},
whose couplings violate lepton number explicitly at a very high scale,
which in turn gives small neutrino masses naturally.
Decays of these triplet higgses generates a lepton asymmetry of the 
universe \cite{ma}. We extend this minimal scenario to include a $SU(2)_S$
gauge group so as to extend the standard model gauge group to 
$${\cal G}_{ext} \equiv 
SU(3)_c \times SU(2)_L \times U(1)_Y \times SU(2)_S. $$ 
The $SU(2)_S$ symmetry breaks down alongwith the lepton number
at very high energy ($M$), and the out-of-equilibrium condition
for generating baryon asymmetry of the universe determines this
scale $M$ \cite{kolb}. Since all representations of the $SU(2)$ 
groups are pseudo-real and anomaly free, there is no additional 
constraints coming from cancellation of anomaly. 
This makes this mechanism easy to implement in different scenarios.

The fermion and the scalar content of the standard model,
which transformations under $SU(3)_c \times SU(2)_L \times U(1)_Y$ as
\begin{eqnarray}
q_{iL} \equiv (3,2,1/6) & u_{iR} \equiv (3,1,2/3)& d_{iR} 
\equiv (3,1,-1/3) \nonumber \\ 
l_{iL} \equiv (1,2,-1/2)& e_{iR} \equiv (1,1,-1)& 
\phi \equiv (1,2,1/2) \nonumber 
\end{eqnarray}
are all singlets under the group $SU(2)_S$.
$i= 1,2,3$ is the generation index. The two heavy triplet
higgs scalars $\xi_a \equiv (1,3,1), 
a=1,2$, required to give masses to the left-handed
neutrinos are also singlets under $SU(2)_S$. In this mechanism 
we add a $SU(2)_S$ doublet neutral left-handed fermion $S_L$ and two scalars
$\eta$ and $\chi$, which transform under ${\cal G}_{ext}$ as
\begin{eqnarray}
S_L \equiv (1,1,0,2) & \eta \equiv (1,2,1/2,2) & \chi \equiv (1,1,0,2).
\nonumber 
\end{eqnarray}

There are two scales in the theory, the  $SU(2)_S$ and the 
lepton number violating scale
$M$ and  the electroweak symmetry  breaking  scale $m_W$.  At  a  high
energy  $M$, $\chi$ acquires  a vacuum  expectation  value ($vev$) and
breaks $SU(2)_S$. Lepton number is broken explicitly at this scale
through the couplings of the scalar triplets.
All the new scalars are considered to be 
very heavy, $$M_\eta \sim M_\chi \sim M_{\xi_a}
\sim  M . $$ The fields $\eta$ and $\xi_a$ do not acquire any
$vev$.  However,  once the  standard   model higgs doublet $\phi$ 
acquires a $vev$, these fields $\xi_a$ and $\eta$ 
acquires a very tiny $vev$, which in turn gives very
small masses and large mixing to the neutrinos. 

Consider the most general potential of all the scalars in the model
($\xi_a, \eta, \chi, \phi$). There will be quadratic and quartic
couplings of the form, $$M_H^2 (H^\dagger H), 
\lambda_H (H^\dagger H)(H^\dagger H)~~{\rm  and}~~\lambda_{12}^\prime
(H_1^\dagger H_1)(H_2^\dagger H_2) ,$$ where $H$ correspond to any one of
the scalar fields. In addition, there will be
two coupled terms,
\begin{eqnarray}
{ V} &=& \mu_a (\xi_a^0 \phi^0 \phi^0 + \sqrt{2} \xi^-_a \phi^+ \phi^0
+ \xi_a^{--} \phi^+ \phi^+) \nonumber \\
&+& m [\phi^0 (\eta^0_{+} \chi_{-}
-\eta^0_{-} \chi_{+}) - \phi^+ ( \eta^-_{+} \chi_{-}
-\eta^-_{-} \chi_{+}) ] \nonumber \\
&+& m [\phi^0 (\eta^0_{+} \chi^*_{-}
-\eta^0_{-} \chi^*_{+}) - \phi^+ ( \eta^-_{+} \chi^*_{-}
-\eta^-_{-} \chi^*_{+}) ] + h.c. 
\end{eqnarray}
where $\eta^-_{+}$ represents the component of $\eta$ with electric
charge $-1$ and $T_3 = +1/2$ of $SU(2)_S$; $\chi_{+}$
represents the component of $\chi$ with $T_3 = +1/2$ of $SU(2)_S$;
and $\chi^*_{+}$
is the component of $\chi^\dagger$ with $T_3 = +1/2$ of $SU(2)_S$.

For consistency \cite{ma} we require $\mu_a$ to be less than
but of the order of masses of $\xi_a$, and we choose 
$$\mu \sim 0.1~M . $$ When the field $\chi$ acquires a $vev$, 
a mixing of the fields $\phi$ and $\eta$ of amount $m <\chi>$ will be 
induced. Since $M_\eta \sim M$ and $m_\phi \sim m_W$, 
to protect the electroweak scale we then require $$m \sim m_W .$$ 
This fixes all the mass parameters in this scenario. We can now 
proceed to minimize the potential. In ref \cite{ma} it was shown 
that the triplet higgs scalars get a very small $vev$ consistent
with the minimisation of the potential. In the present scenario
both the higgs triplet $\xi_a$ and the new doublet higgs scalar $\eta$
get a tiny $vev$ on minimisation, without any fine tuning of 
parameters. We assume that $T_3 = +1/2$ component of $\chi$ acquires
a $vev$. But that can induce $vev$s to both the neutral 
$SU(2)_S$ components $\eta_+^0$ and $\eta_-^0$. They are given by,
\begin{eqnarray}
<\xi_a> &\simeq& - {\mu <\phi>^2 \over M_{\xi_a}^2} \nonumber \\
<\eta^0_-> \simeq - {m <\phi> <\chi_+> \over M_{\eta}^2} &{\rm and}&  
<\eta^0_+> \simeq - {m <\phi> <\chi_-^*> \over M_{\eta}^2} .
\end{eqnarray}
Since, $\mu \sim M_{\xi_a} \sim M_{\eta} \sim <\chi_+> \sim M$
and $m \sim m_\phi \sim <\phi> \sim m_w$, we get, 
\begin{equation}
<\xi_a>~~\sim ~~<\eta_-^0>~~ \sim  ~~<\eta^0_+>~~\sim~~
O \left( {m_W^2 \over M} \right) .
\end{equation} 
The $vev$s of $\xi_a$
now give small masses to the left-handed neutrinos and the $vev$ of 
$\eta^0_\pm$ allows mixing of the $SU(2)_L$ singlet neutrinos $S_L$ with the
usual left-handed neutrinos, both of which are now of the same order of 
magnitude $\underline{naturally}$. 

The Yukawa couplings of the leptons are given by,
\begin{equation}
{\cal L} = f^e_{\alpha i} \overline{l_{iL}} e_{\alpha R} \phi 
+ f_{aij} l_{iL} l_{jL} \xi_a + h_{ix} \epsilon_{xy} l_{i L} S_{Lx} \eta_y 
+ h.c.  \label{yuk1}
\end{equation}
where $x,y = 1,2$ are the $SU(2)_S$ indices. The first term contributes
to the charged lepton masses, while the second and third terms contributes
to the neutrino mass and mixing matrix. In the basis, $[\nu_{i L}~~S_{Lx}]$
we can now write down the mass matrix as,
\begin{equation}
{\cal M}_\nu = \pmatrix{ \sum_a f_{aij} <\xi_a> & h_{ix} 
\epsilon_{xy} <\eta_y^0> \cr
h^T_{ix} \epsilon_{xy} <\eta_y^0> & 0} . \label{mass}
\end{equation} 
There is no Majorana mass terms for the sterile neutrinos. We shall
now discuss how to generate baryon asymmetry of the universe 
\cite{sakh,kolb} in this
scenario and what constraint it gives on the new mass scale $M$
and then come back to the neutrino mass matrix.

Lepton number is violated when the scalars $\xi_a$ decays 
\begin{equation}
\xi_a \rightarrow \left\{ \begin{array} {l@{\quad}l} l_i^c l_j^c & 
(L = -2) \\ \phi \phi & (L = 0) \end{array} \right.
\end{equation}
All other couplings conserve lepton number. By assigning a lepton number
$-1$ to $S_{Lx}$, we can ensure conservation of lepton number in the
decays of the doublet scalar field $\eta$. 

We choose the mass matrix of $\xi_a$ to be real
and diagonal $\pmatrix{M_{\xi_1} & 0 \cr 0 & M_{\xi_2}}$; but once 
the one loop self energy type contributions are included, imaginary phases
from $\mu_a$ and $f_{aij}$ makes it complex. The absorptive part
of the one loop self-energy type diagram will introduce observable 
CP violation in the mass matrix \cite{paschos}, which would produce
unequal amount of leptons and anti-leptons in the decays of the 
$\xi_a^{++}$ and $\xi_a^{--}$ respectively. This will create a charge
asymmetry, which will be compensated by equal and opposite amount
of charge asymmetry in the production of $\phi^+$ and $\phi^-$ in the
decays of $\xi_a^{++}$ and $\xi_a^{--}$, so that the universe remains
charge neutral.

The interference of the tree level
and the one loop diagram of figure 1 will generate a lepton asymmetry in the
decays of $\xi_a$, which is given by,
\begin{figure}[t]
\vskip 2.5in\relax\noindent\hskip -.5in\relax{\includegraphics{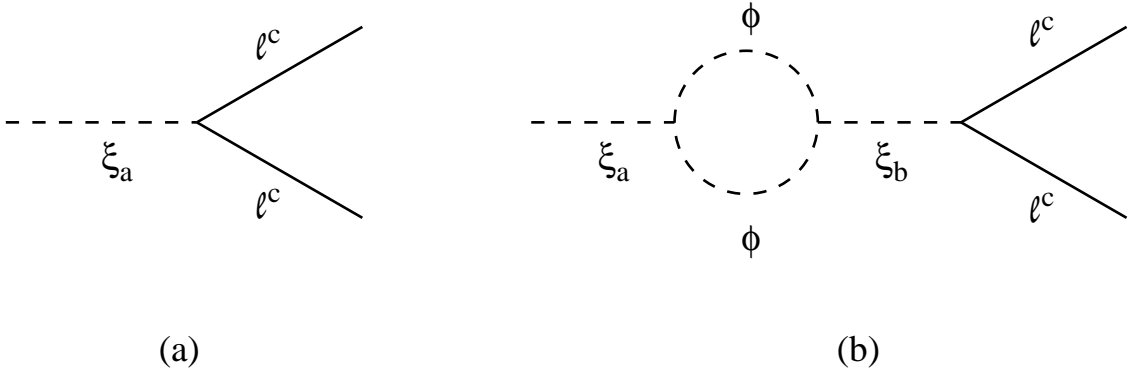}}
\caption{The decay of $\xi_a \to l^c l^c$ at tree level (a) and in 
one-loop order (b).  $CP-$violation comes from an interference of these 
diagrams.}
\end{figure}
\begin{equation}
\delta_a \simeq 
{{Im \left[ \mu_1 \mu_2^* \sum_{k,l} f_{1kl} f_{2kl}^* \right]} \over 
{8 \pi^2 (M_{\xi_1}^2 - M_{\xi_2}^2)}} \left[ {{ M_{\xi_a}} \over 
\Gamma_{\xi_a}}  \right].
\end{equation}
where, the decay width of these scalars $\xi_a$ is given by,
\begin{equation}
\Gamma_{\xi_a} = {\frac{1 }{8 \pi}} \left( {\frac{|\mu_1|^2 + |\mu_2|^2
}{M_{\xi_a}}} + \sum_{i,j} |f_{aij}|^2 M_a \right) .
\end{equation}
These decays should be slower \cite{kolb} than the expansion rate of the 
universe ($H$), otherwise the lepton asymmetry $\delta_a$ in decays
of $\xi_a$ will be suppressed by an amount $K ~({\rm ln}~K)^{0.6}$, where
$K = {\Gamma_{\xi_a} \over H}$; $
H = \sqrt{1.7g_*}\frac{T^2}{M_{Pl}} ~~~{\rm at}~~T=M_{\xi_a}$; 
$M_{Pl}$ is the Planck scale; and 
$g_*$ is the total number of relativistic degrees of freedom.

We consider $M_{\xi_2} < M_{\xi_1}$, so that when $\xi_2$ decays, $\xi_1$
has already decayed away and only the asymmetry $\delta_2$ generated in
decays of $\xi_2$ will contribute to the final lepton asymmetry of the 
universe. 
The lepton asymmetry thus generated will be the same as the $(B-L)$
asymmetry of the universe, which will then get converted to a 
baryon asymmetry during the electroweak phase transition \cite{ht}.
The final baryon asymmetry of the universe is given by,
\begin{equation}
{n_B \over s} \sim {\delta_2 \over 3 g_* K ({\rm ln}~K)^{0.6}} .
\end{equation}
To obtain the desired amount of baryon asymmetry of the universe
one possibility could be \cite{ma}, 
$M_2 = 10^{13}$ GeV, and $\mu_2 = 
2 \times 10^{12}$ GeV, which gives us $m_{\nu_\tau} = 1.2 f_{233}$ eV, 
assuming that the $M_1$ contribution is negligible.  Now let $M_1 = 3 \times 
10^{13}$ GeV, $\mu_1 = 10^{13}$ GeV, and $f_{1kl} \sim 0.1$, then the decay of 
$\psi_2$ generates a lepton asymmetry $\delta_2$ of about $8 \times 
10^{-4}$ if the CP phase is maximum.  Using $M_{Pl} \sim 10^{19}$ GeV and $g_* 
\sim 10^2$, we find $K \sim 2.4 \times 10^{3}$, so that $n_B/s 
\sim 10^{-10}$.

Thus, with the heavy mass scale to be of the order of $M \sim
10^{13-14}$ GeV it is possible to get the desired amount of baryon 
asymmetry of the universe and $vev$s of $\xi$ and $\eta$ to be of the
order of a few eV. Then with proper value of the Yukawa couplings 
$f_{aij}$ and $h_{ix}$ we
can get a neutrino mass matrix (equation \ref{mass})
which can explain all the neutrino 
experiments. All the elements of the mass matrix could be non-zero
and are about a few eV or less, except
for the Majorana mass term of the sterile neutrinos. 
One can then have several possible scenarios \cite{rev}
which can explain all
the neutrino experiments. Cosider for example \cite{mass}
one sterile neutrino ($S_{L1}$) 
with mass of about $(2-3)\times 10^{-3}$ eV, which mixes with 
the $\nu_e$, while the other sterile neutrino ($S_{L2}$) is lighter.
This will satisfy the constraints on the sterile neutrinos from 
nucleosynthesis. If we choose $m_{\nu_e}$ to be much lighter than $S_{L1}$,
that satisfies all laboratory constraints on its mass. Then 
the $\nu_e \to \nu_{S_1}$ oscillations can explain the solar neutrino
deficit. We can further assume that $\nu_e \to \nu_{S_1}$ oscillations 
satisfies the resonant oscillation condition, while $\nu_e \to \nu_{S_2}$ 
oscillations satisfies the vacuum oscillation condition, so there are 
both the components. 

For a solution of the atmospheric neutrino anomaly we assume 
the $[\nu_\mu ~~\nu_\tau]$ mass matrix to be of the form,
\begin{equation}
M = \pmatrix{ m_a & m_{ab} \cr m_{ab} & m_b}
\end{equation}
where, $m_{ab} > m_b > m_a$, so that the two physical states are almost
degenerate with masses $m_{ab}$, but the mass squared difference 
is given approximately by, $m_{\nu_\mu}^2 - m_{\nu_\tau}^2 \sim 
m_{ab} m_b$. We can then have $m_{ab} \sim $eV and $m_b \sim 10^{-2}$ eV,
so that the $\nu_\mu$ and $ \nu_\tau$ are almost degenerate with
mass about a few eV to be the hot component of the dark matter and
the mass squared difference $m_{\nu_\mu}^2 - m_{\nu_\tau}^2 \sim
10^{-2}$ eV$^2$ and maximal mixing can explain the atmospheric 
neutrino anomaly. These numbers show the freedom available to the present 
scenario to explain all the data. In practice, as in the case of 
quark and charged lepton masses, only future experiments can 
determine the exact form of the Yukawa couplings $f_{aij}$ and $h_{ix}$. 

To summarise, we propose a simple scenario to accomodate naturally
light sterile neutrinos in extensions of the standard model with an
$SU(2)_S$ symmetry. There are two mass scales in the model, the 
electroweak scale and the scale of lepton number and $SU(2)_S$
violation, which is
fixed by the conditions for lepton asymmetry of the universe. The 
heavy triplet and a doublet acquires very tiny $vev$, which gives 
masses and mixing of the left-handed neutrinos and the sterile neutrinos.
The low energy mass matrix can then explain the solar neutrino deficit,
atmospheric neutrino anomaly, LSND result and the dark matter problem.
The decays of the triplets generates a lepton asymmetry of the universe, 
which gets converted to a baryona symmetry of the universe during the 
electroweak phase transition. 

\vskip 0.5in
\begin{center} {ACKNOWLEDGEMENT}
\end{center}

I would like to thank Prof W. Buchmuller and the Theory Division, DESY,
Hamburg for hospitality and acknowledge financial support from the
Alexander von Humboldt Foundation.

\newpage
\baselineskip 18pt
\bibliographystyle{unsrt}

\end{document}